\documentstyle[11pt,aaspp4]{article}
\def\um {$\mu$m\ }

\doublespace
\begin{document}
\slugcomment{Resubmitted RSI A02299, 2002 Jun 27}
\title{A LOW NOISE THERMOMETER READOUT FOR RUTHENIUM OXIDE RESISTORS}
\begin{center}
\author{D. J. Fixsen, P. G. A. Mirel}
SSAI code 685\\
NASA/GSFC Laboratory for Astronomy and Solar Physics\\
Greenbelt MD 20771
\author{A. Kogut}
NASA/Goddard Space Flight Center\\
Laboratory for Astronomy and Solar Physics\\
Greenbelt MD 20771
\author{and M. Seiffert}
Jet Propulsion Laboratory, California Institute of Technology\\
4800 Oak Grove Drive\\
Pasadena, CA 91109
\end{center}
\begin{abstract}
The thermometer and thermal control system, for the Absolute Radiometer for 
Cosmology, Astrophysics, and Diffuse Emission (ARCADE) experiment, is described,
including the design, testing, and results from the first flight of ARCADE.
The noise is equivalent to about 1~$\Omega$ or 0.15 mK in a second for the 
RuO$_2$ resistive thermometers at 2.7~K. The average power dissipation in each 
thermometer is 1~nW. The control system can take full advantage of the 
thermometers to maintain stable temperatures. Systematic effects are still 
under investigation, but the measured precision and accuracy are sufficient to 
allow measurement of the cosmic background spectrum.

\end{abstract}
\clearpage
\section{Introduction}
The Absolute Radiometer for Cosmology, Astrophysics, and Diffuse Emission 
(ARCADE) experiment$^1$ is designed to detect or limit spectral 
distortions in the Rayleigh-Jeans tail of the Cosmic Microwave Background.
The key to this experiment is the external calibrator. The absolute 
temperature of the calibrator is required to compare to other experiments
such as Far InfraRed Absolute Spectrophotometer (FIRAS)$^{2,3}$.
It is more critical, however, that
the calibrator be isothermal and remain at a constant temperature while
it is shifted between the various radiometers of ARCADE. The other parts
of the ARCADE instrument (loads, horns, switches etc) must remain at a 
stable temperature, but the absolute accuracy requirement is not severe
because the external calibrator will calibrate all of these terms to 
first order. The measurement and thermal control must be performed at
high altitudes ($\sim30$~km) while the instrument is suspended from a balloon.

In order to eliminate the reflections and emission of a window the ARCADE 
radiometers run without a window. The target, similar in many respects to that
of FIRAS, is moved from one horn to the other to provide an external blackbody
reference to compare to the sky. A blackbody internal reference reduces
the dynamic range of the amplifier signal.
With this arrangement, it is important to maintain the temperature of critical
components in the radiometer such as the target, the internal reference, 
the amplifier and the horn antenna while the target is moved from one
radiometer to another.

Cryogenic thermometers are often used in applications where low noise is
desired and low power is required. Low level signals from the thermometers
are susceptible to noise pick-up and degradation by the capacitance of long
lines. Long term measurement stability is required, so it is desirable to 
have a thermometer system with built in calibration.

Most of the ARCADE thermometers are used to maintain thermal control of the 
instrument; low noise and stability are more important than knowledge of the 
absolute temperature. However, key thermometers are imbedded within the 
microwave absorber of the external calibration target. The ARCADE science 
goals require an isothermal target, which in turn requires precise 
cross-calibration for the target thermometers. 

\section{THERMOMETER DESIGN}
The system requirements include about 25 thermometers, with a $\sim$1 Hz read
rate, and low power dissipation, with four wire measurements. Drifts in the 
readout system are a concern for balloon operations so fixed resistors are
included for self calibration.

RuO$_2$ resistive thermometers were chosen because of their low cost, long term
stability and large resistance changes in the neighborhood of 2.7 K$^4$.
The thermometers are commercially-available thick-film 
chip resistors$^5$ (Fig~\ref{Picture}), with resistance 10~k$\Omega$ 
at room temperature. At 2~K, the resistance rises to $\sim$40~k$\Omega$ with a 
$dT/dR$ of -0.098, -0.158, -1.042, and -10~mK/$\Omega$ at 2.2, 2.7, 6.3 and 
20~K respectively. The readout noise is ultimately related to this ratio.
Manganin leads (76~\um\ dia.) allow 
4-wire resistance measurements. Mechanical stress is relieved by thermally 
cycling each chip 100 times between 300 K and 77~K prior to soldering the 
leads, then another 50 times after the leads are attached. Most of the 
thermometers monitor the cryogenic performance of the instrument 
(e.g., amplifier or dewar temperatures). The thermometers are mounted
to a thin copper tab using Stycast 2850-FT epoxy, and calibrated against
National Institute of Standards and Technology (NIST)-traceable Ge 
thermometers on the cold stage of a pumped LHe dewar.

The entire target, including five embedded thermometers, was
cycled an additional ten times to relieve thermal stress, then the thermometers 
were calibrated {\it in situ} by immersing the entire target in a liquid helium
bath and varying the vapor pressure above the bath. Temperature gradients
within the target during calibration are small, limited ultimately by
convection cells within the liquid. The absolute temperature scale
is referenced to a NIST-traceable Ge thermometer and cross-checked by
measurement of the superfluid helium transition temperature. At temperatures
near 2.7 K, the target thermometer calibration absolute accuracy is 
$\sim$0.9 mK verified by repeated cycling and remeasurement.

The readout utilizes a four-wire resistance measurement, with an alternating 
excitation current and a lock-in integrator for low-power and low-noise$^6$.
To minimize the power dissipated in the thermometer, the current is limited to
1.3~$\mu$A. The readout multiplexes among 28 thermometers; the lines to
each thermometer vary in length, so the line capacitance seen by the readout
varies. To make each measurement in 1/30 second and still use an alternating 
current to minimize the effects of offset drifts, 75~Hz was used for
the excitation frequency. The circuit is shown in Figure~\ref{Circuit}. Note 
all of the components except for the thermometers are maintained in a RF 
shielded box with a controlled temperature.

The readout of each thermometer is accomplished in 5 phases. Each phase is
6.7~ms long. During phases 1 \& 3 the excitation current and integration
signal are positive, while for phases 2 \& 4 both are negative 
(Fig~\ref{Phases}). During the fifth phase the current is zero.
During the first half of the fifth phase the integrator is decoupled
from its input and the result is digitized. During the second half
of the fifth phase the integrating capacitor is shorted to reset it for
the next thermometer. At the transition from the phase 5 to phase 1
the current and signal multiplexers are switched to the next thermometer
(or to a reference resistor).
The readout multiplexes among four calibration resistors as well as the 28
thermometers. The 32 inputs are read sequentially, and the results are
sequentially relayed to a computer with RS232 protocol at 2400 Baud.
The multiplexing and the off phase together reduce the duty cycle to
2.5\%, so the average power dissipation is only $\sim1$~nW.
The same clock is used for the instrumentation phases as for the digital 
readout, so the time for a single sample is 1/30 second. Thus it takes 
1.067 seconds for a complete cycle.

The integrating capacitor is polypropylene for stability (Sprague 730P).
The charging time constant is 20~ms. Low resistance discharge resistors make 
the discharge time constant 0.2~ms. While the signal is digitized the
capacitor is essentially disconnected from any input. Figure~\ref{Timing} 
shows the signal on the integrating
capacitor for five samples during a laboratory test. The digitize command
(also shown) initiates the digitization at the beginning of phase 5 of the
cycle just as the integrating capacitor is decoupled from the input and the
current to the thermometer is zeroed.

The metal film calibration resistors were characterized: The observed
change in their resistance is less than 0.5\% between 295~K and
77~K. The values of these resistors are distributed over the dynamic
range of the RuO$_2$ thermometers. 

\section{PRINTED CIRCUIT BOARD DESIGN}

To minimize the noise pickup between components, the printed circuit board 
was designed with internal ground and power planes. The circuit traces were 
distributed on the top and bottom of the board to allow for field modification 
of the circuit. The analog components were clustered around the opamps to keep 
the leads short, and the digital components were pushed to the outer edges of 
the circuit board to minimize their coupling into the measurement circuitry. 
The multiplexer chips were kept close to the input connector
and the opamp to minimize line length.

\section{TESTING DESCRIPTION}

During testing, the circuitry and wiring were optimized for low noise. The
offset resistor, nominally 400~k$\Omega$, was tuned to 329~k$\Omega$ to null 
the integrator reading at 25~k$\Omega$ (midscale or 2.7~K). 
Capacitance (50 pF) was added in parallel with each calibration resistor to 
approximately match the measured line capacitance to the sensors, and 
30~$\Omega$ was added in series on each end to match the line resistance. The 
feedback capacitance in the excitation stage was set at 1 nF to roll off of the 
square wave to reduce the high frequency noise while maintaining 
sufficient fidelity in the square wave to avoid spikes in the demodulation.
The output time constant is 0.2 ms. The cables connecting 
the instrument box to the dewar are 41 conductor braid-shielded cable. The 
shielding was grounded at both ends to minimize ground pickup noise, and the 
thermometer cables were separated from the cables that carry the
driver lines to the ferrite switches in the instrument.

Nominally the readout voltage-to-ohm conversion is linear. For flight
instrumentation and the temperature control (see \S5), a simple linear
fit between the 20~k$\Omega$ and 25~k$\Omega$ reference resistors
was used. For detailed post-flight analysis calculations a second order
fit was used for all four of the reference resistors. The data for the 
reference resistors were smoothed over 32 second intervals to minimize
the noise contribution from the reference resistors. The second order fit
only changes the value by a few ohms (about 1~mK).

Once the resistance was determined, the table of calibration temperatures,
and resistances for each resistor, obtained in ground calibration testing,
was interpolated to arrive at a temperature estimate. The final uncertainty is 
dominated by the calibration process rather than the readout uncertainty.

\section{THERMAL CONTROL}
Stable temperatures of the radiometer components are required for the 
ARCADE science mission. To determine the coupling parameters of the 
radiometer to individual components, it is important to be able to change the 
temperature of those individual components selectively. To accomplish this, the 
components were placed under closed-loop thermal control.

Closed-loop thermal control was performed using; 1. the temperature sensing 
described above, 2. a simple resistive heater driven by a D$\rightarrow$A, 
and 3. a software proportional-integral-differential (PID) control loop. A
PID control loop imposes a restoring action proportional to the error signal
(the P). In many situations this leads to oscillation so a damping term is
added (the D). But an external bias leads to a long term error so an integral
correction is included (the I).The signal for the control loop, in this case, 
was the difference between the measured temperature and the desired 
temperature (in K). The control output is calculated in watts and then 
converted to the appropriate value for the D$\rightarrow$A and heater driving 
circuits. Ten control loops were implemented for the various 
instrument components. 

The PID loop has an update rate of $\sim$1~Hz, limited
by the update rate on the temperature sensors.  The PID parameters
for each item to be controlled were determined empirically by using 
measured time constants and oscillation frequencies. The differential
parameter was intentionally set to a value closer to zero than would
be optimal for transient response.  This allowed us to have 
robust control loops that do not oscillate over
a wide temperature range, at a small cost in settling time after
a commanded temperature change.

For each loop, only a single set of parameters was used to cover the 
temperature range from 2 - 20 K. Typical time constants for internal 
instrument components were of order 10 seconds.
Typical control parameters were of order $k_p = 0.1$~(W/K) (proportional
term), $k_i = 0.015$~(W/K-s) (integral term), and $k_d = -0.02$~(W-s/K) 
(differential term).


If the temperature error was more than a maximum value (typically
5 K), the PID loop was temporarily suspended and the heater turned full
on or off. Once the temperature error was reduced
to below the maximum value, PID control was restored. This loop override
was useful for making quick large temperature changes.
Figure~\ref{Control} shows the in-flight thermal control
performance of the 10 GHz internal load. 

\section{PERFORMANCE IN FLIGHT}
Performing temperature measurement at $\sim$30~km altitude puts reliability and
simplicity at a premium. Care must be taken to avoid arcing (far easier
at low pressure), and to make the circuit robust in the face of uncertain
temperatures and launch shocks. However, the battery powered gondola has
advantages as well. The circuit is far away from the 60~Hz that plagues
laboratory measurements. And the balloon ride is very smooth (after launch)
eliminating the ground vibration. Even the air currents are light (as the
balloon follows the wind), stable and at low pressure. These together allow
the inflight noise to be lower than the best measured noise in the laboratory
before launch.

The 4 reference resistor measurements drifted only by about five counts 
(4~$\Omega$) over the entire flight. Most of the changes occurred in the form 
of discrete steps or jumps. Data near the jumps was excised. 
The cause of these steps was not determined although they appeared in
the measurements of all of the thermometers as well. After calibration the RMS
noise of the measurements was approximately 1~$\Omega$ for each sample 
(Fig~\ref{Ref_Noise}). This corresponds to 0.15~mK on the RuO$_2$ thermometers 
at the critical temperature of 2.7~K, comparable to bridge measurements
done in the laboratory$^7$.

The final performance is shown in Figure ~\ref{Noise}. It is not clear if 
the low frequency
noise (below 0.02~Hz) is low frequency drift in the thermometry system or
actual drifts in the temperature of the instrumented component. However by 
shifting the target between the different frequency antennas every minute or 
so, the low frequency noise will not be an issue in measuring the Cosmic 
Background Radiation at the sub mK level.

\section{Acknowledgement}
We would like to thank the IR laboratory staff of the NASA/GSFC Code
685. We are also grateful to Leah Johnson, Elizabeth Cantando, and Paul 
Alexandre Rischard for assisting with the testing and installation during their 
summer program, and Anatoly Brekhman and Gary Palmer II for wiring the 
thermometers during their school-year internships.

\clearpage
\typeout{FIGURE CAPTIONS}
\begin{figure}
\caption[RuO$_2$ surface mount resistor]
{\protect\small
\protect\addtolength{\baselineskip}{-0.5ex}
A RuO$_2$ surface mount resistor is shown.  The marks on the scale are .01 inches.
\label{Picture}}
\end{figure}

\begin{figure}
\caption[Thermometer Readout Circuit]
{\protect\small
\protect\addtolength{\baselineskip}{-0.5ex}
Circuit diagram shows the major elements of the RuO$_2$ readout circuit.
All of the multiplexers are CD4052 dual 4-to-1 analog multiplexers. The 
opamps are
an LT1125 quad opamp. The integration capacitor is polypropylene for stability.
\label{Circuit}}
\end{figure}

\begin{figure}
\caption[Phase Diagram]
{\protect\small
\protect\addtolength{\baselineskip}{-0.5ex}
Shown here is the oscilloscope trace of the excitation current. The 
A$\rightarrow$D converter is activated at the beginning of the fifth phase.
\label{Phases}}
\end{figure}

\begin{figure}
\caption[Timing Diagram]
{\protect\small
\protect\addtolength{\baselineskip}{-0.5ex}
Shown here is the oscilloscope trace of five samples. The first sample is near 
33~k$\Omega$. The others are: 25~k$\Omega$ (the null point), 20~k$\Omega$, open, 
and 10~k$\Omega$ respectively. Also shown is the digitize command strobe.
\label{Timing}}
\end{figure}

\begin{figure}
\caption[Temperature Switching]
{\protect\small
\protect\addtolength{\baselineskip}{-0.5ex}
Here is 270 seconds of temperature data, demonstrating the control in 
switching between 6.2 and 6.4~K.  The noise is higher at 6~K than at 2.7~K
because of the $dR/d$T dependance on T. A careful look at the transitions
shows a $\sim$20 settling time.
\label{Control}}
\end{figure}

\begin{figure}
\caption[In Flight Noise Measurement]
{\protect\small
\protect\addtolength{\baselineskip}{-0.5ex}
Noise spectrum of one of the calibration resistors. This spectrum
was taken between the relatively rare jumps of $\sim$5 counts.
Each count corresponds to .8~$\Omega$.
\label{Ref_Noise}}
\end{figure}

\begin{figure}
\caption[In Flight Performance]
{\protect\small
\protect\addtolength{\baselineskip}{-0.5ex}
Noise spectrum of one of the thermometers at 2.7~K.
\label{Noise}}
\end{figure}

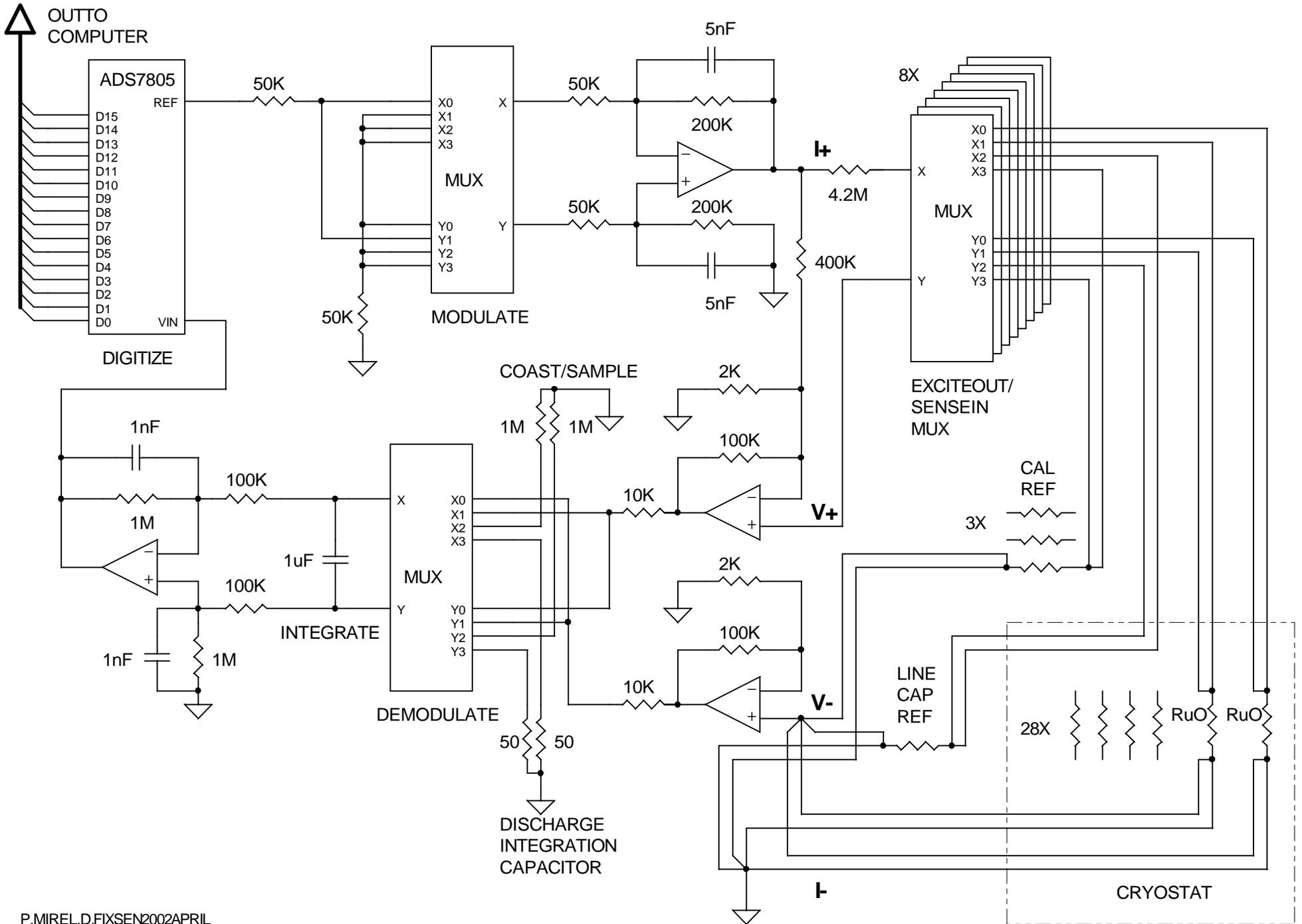
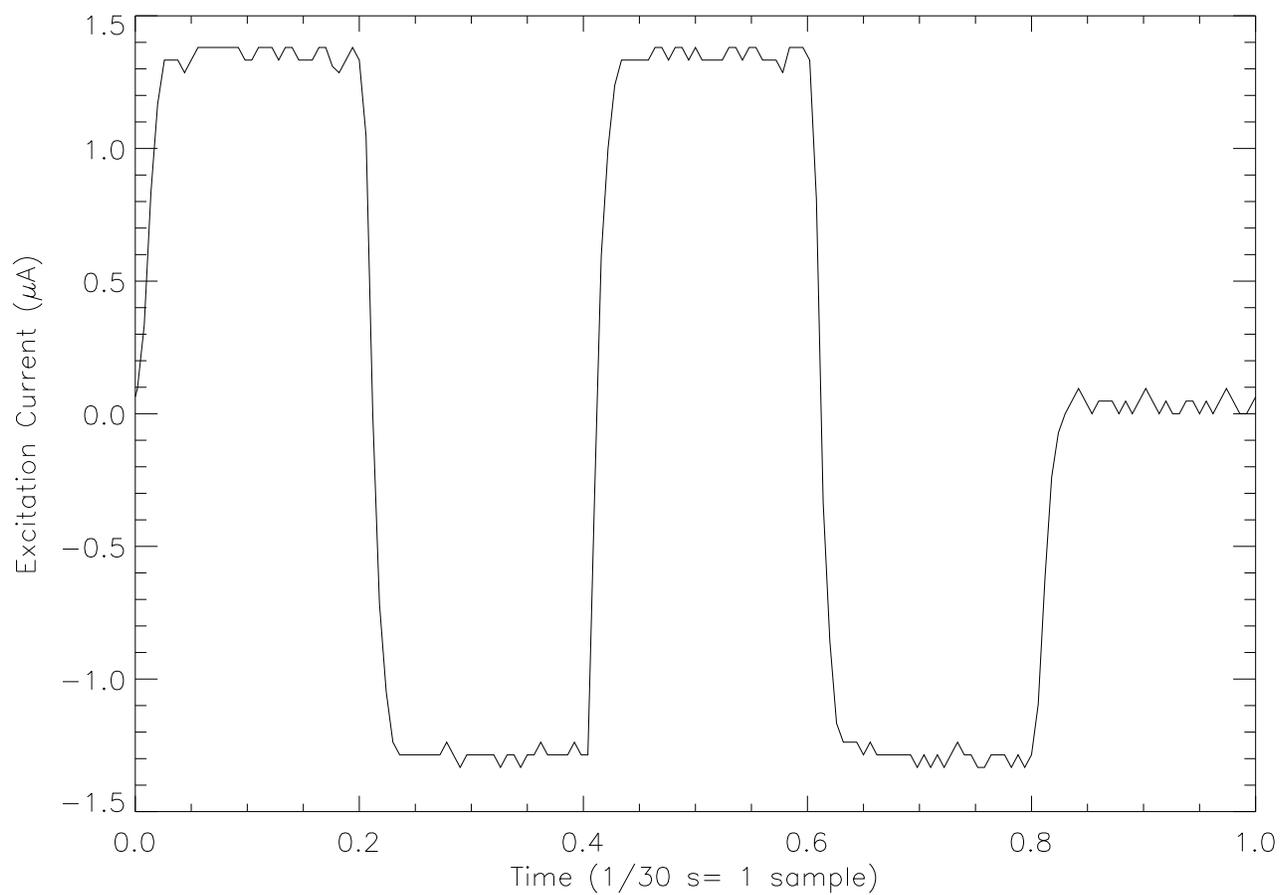
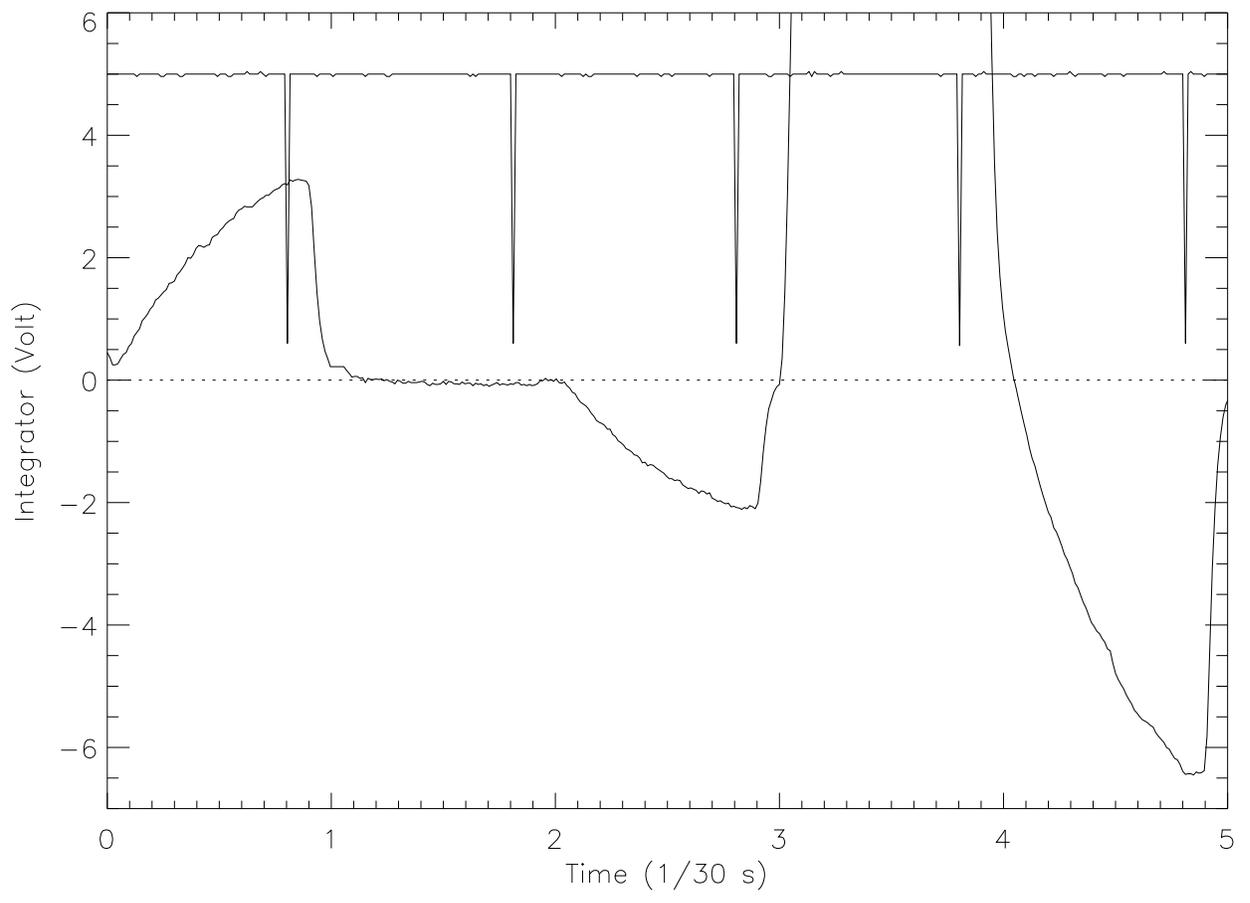
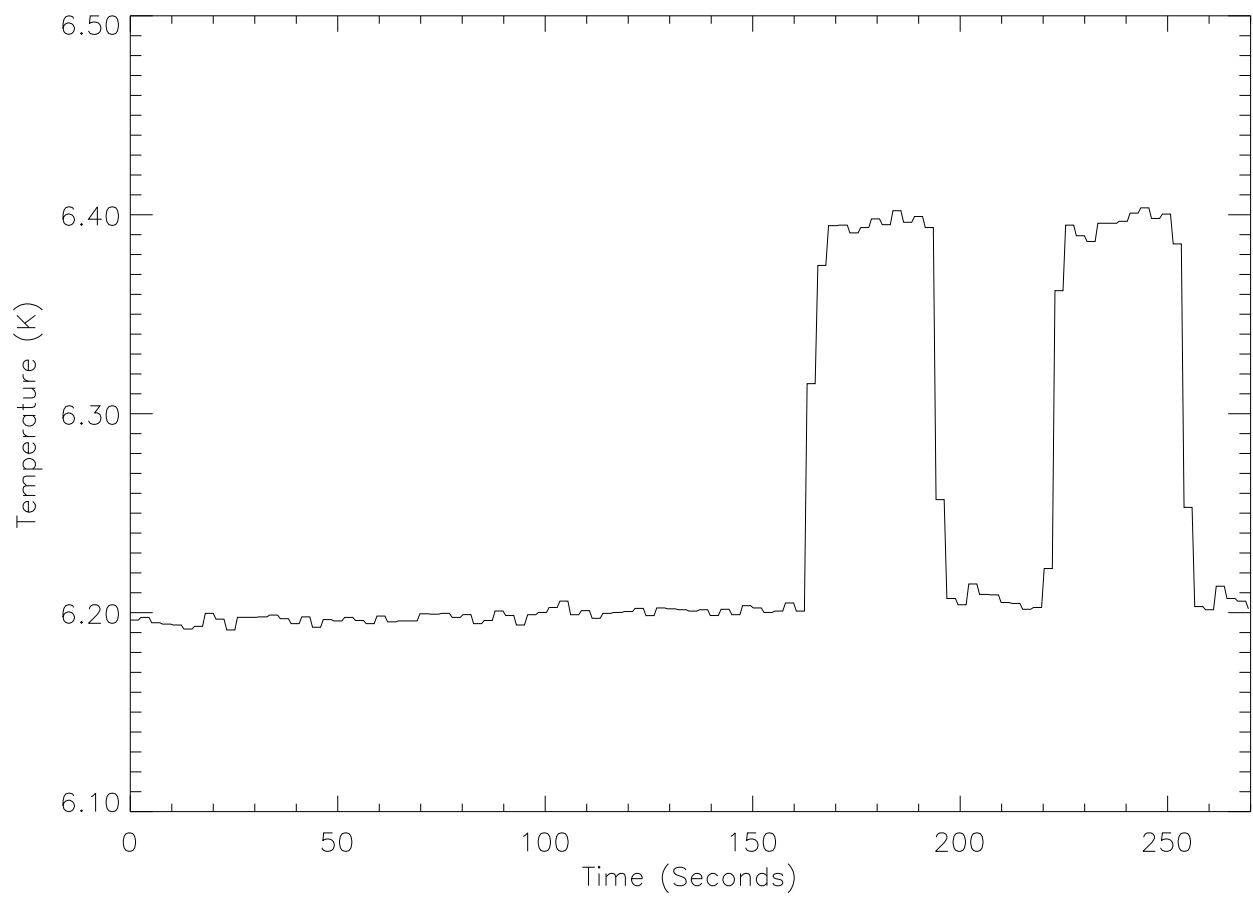
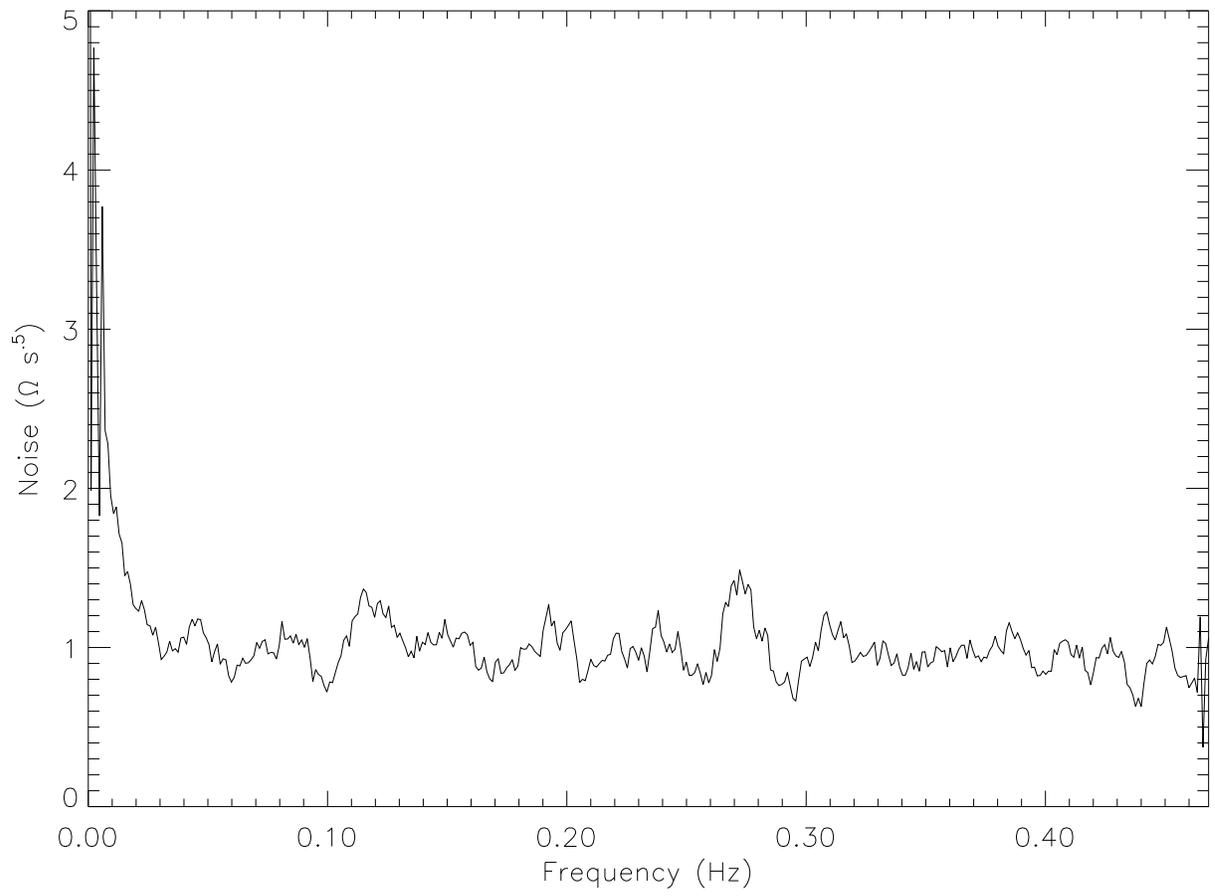
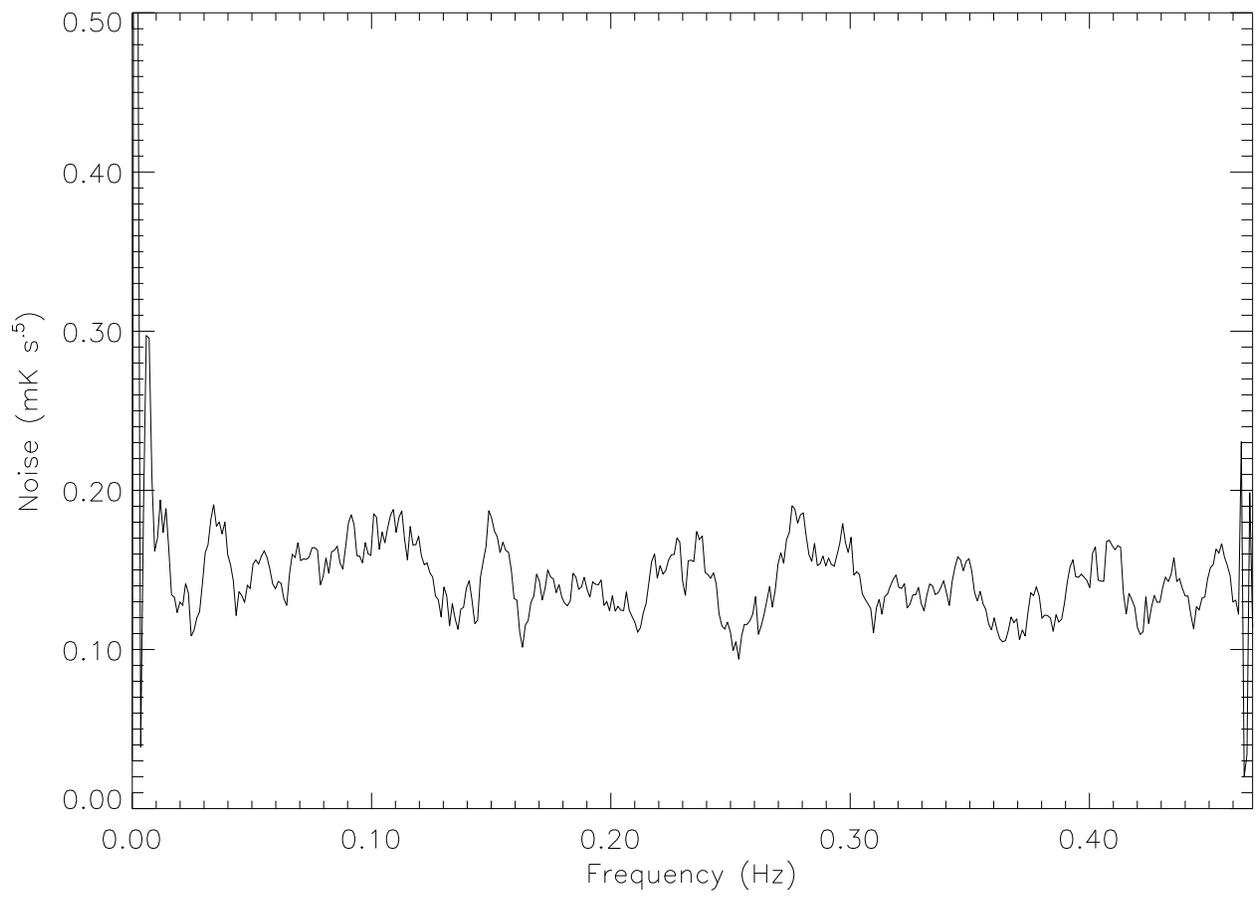
\end{document}